\documentclass{iau}

\usepackage{amsmath}
\usepackage{graphicx}
\usepackage{multirow}
\usepackage{aas_macros}
\usepackage[T1]{fontenc}

\begin{document}

\lefttitle{Apai, Lin, \& Wagner}
\righttitle{The Interstellar Laser Beacons Hypothesis}

\jnlPage{1}{7}
\jnlDoiYr{2021}
\doival{10.1017/xxxxx}

\aopheadtitle{Proceedings IAU Symposium}
\editors{J. Haqq-Misra \&  R. Kopparapu, eds.}

\newcommand{\hypothesis}{Interstellar Beacons Hypothesis}

\title{The Interstellar Laser Beacons Hypothesis and the Cosmic Lighthouses Project}


\author{D\'aniel Apai, Chia-Lung Lin, Kevin Wagner}
\affiliation{Steward Observatory and Lunar and Planetary Laboratory,The University of Arizona}
\affiliation{James C. Wyant College of Optical Sciences,The University of Arizona}
\affiliation{Earth, Atmospheric, and Planetary Sciences, MIT}
\begin{abstract}
In this paper, we argue that in a realistic, resource-limited scenario senders whose intent is to maximize successful communication at the lowest cost will utilize very long-lived, pulsed beacons. Further, we argue that mass-producible laser beacons with simpler technology ($\mu$s or ms pulses instead of ultrashort fs pulses) are more optimal if placed on interstellar orbits. Our extensive, existing surveys are mostly blind to the $\mu$s/ms-second universe and to such beacons. We show, however, that the convergence of three technologies (massively parallel processing, ultra high-speed CMOS detectors, and broadband, mass-producible diffractive-refractive lenses) now offer an exciting opportunity for a low-cost survey of the $\mu$second sky. Such survey would be astrophysically compelling and, with even low-cost telescopes, could place volume-complete constraints on interstellar beacons within twenty parsecs.  
\end{abstract}


\maketitle




\section{Communication in a Resource-Limited World}

Resource constraints can alter ideal strategies for interstellar communication \citep[][]{Shostak2009,Benford2010}. Here, we investigate such considerations and find arguments that suggest that low-cost, resilient, and abundant interstellar laser beacons may be a key component of communications between extraterrestrial civilizations. 
In this paper, we make two assumptions: (a) The sender's presumed goal, by sending a message, is to influence the receivers' behavior in some way, i.e., to reach an intended action; (b) The sender is resource-limited and thus optimizes their project to minimize cost while maximizing likelihood of success. 

Arguably, the probability of reaching the goal of an intended action from the side of receiver can be expressed as a combined probability of a sequence of events:
\begin{equation}
P_{success} = P_{det} \times P_{ident} \times P_{interp} \times P_{intended\,action},
\end{equation}
where the terms on the right describe the probability of successful detection of signal ($P_{det}$), the identification of the data as a signal ($P_{ident}$), successful interpretation ($P_{interp}$), and the probability that it compels the receiver to undertake the intended action $P_{intended\,action}$.

We simply represent the cost of the communication attempt as
$C \propto C_R \times t_d + C_{NR},$

where $C_R$ represents recurring costs (or ``operational expenses'') per unit time, $t_d$ is the duration of the project in the same units. $C_R$ could include transmission cost, maintenance and unit replacement, and other operational costs. $C_{NR}$ represents non-recurring costs (or ``capital expenses''). These may include technology development, infrastructure installation, launch costs,  formulation of the message, etc. 

The above simple model illustrates that the sender is compelled to craft a message that is clear and will lead to the intended result, i.e., maximizing  $P_{interp} \times P_{intended\,action}$. This is a  non-recurring, low-cost task. In contrast, maximizing $P_{det}$ is likely a recurring task (continued emission) that requires a capable transmission infrastructure.
The design, construction, deployment and operation of that transmission infrastructure is the likely cost driver for both the recurring $C_R$ and non-recurring costs $C_{NR}$.   

We also note, that due to the immense search space and light-travel times, serious attempts for interstellar communication likely plan for projects lasting centuries if not millennia. This is an important consideration as it incentivizes the sender to use highly automated, extremely robust equipment. It is likely that robust equipment will not be highly complex and finely-tuned systems that are pushing the limits of what is physically possible (e.g., high-intensity femto-second lasers) but may use scalable systems that consists of large numbers of simpler, robust, mass-producible unit (e.g., arrays of $\mu$s lasers). 





\section{Interstellar Laser Beacons}

The high-dimensionality of the parameter space for interstellar communication makes communication challenging. Signal localization (in space and in time) collapses four dimensions into a single point and thereby vastly increases the probability of success. A resource-limited sender will be compelled to aid localization while keeping costs down. It has been argued that beacons are a powerful option for signal localization \citep[e.g.,][]{Benford2010} and pulsed beacons may be optimal \citep[]{Shostak2009}. 

In this work, we consider laser beacons as a potentially compelling option for resource-constrained senders. There are good reasons for considering light as a medium for communication: Arguably, signals emitted at wavelengths close to peak of their host star's emission are likely detectable by most civilizations and likely even without advanced technology. Similarly, artificial coherent emission at visible or near-infrared wavelengths is not a very challenging technology: In our civilization, lasers were  invented \citep[][]{Maiman1960} relatively shortly after radio telescopes \citep[e.g.,][]{Jansky1933}. The modulation of laser emission is straightforward and very short pulses can pack high power and high information content. Their potential for interstellar messaging was realized early \citep[][]{Schwartz1961} and continues to be a viable option \citep[][]{Shostak2009}. Furthermore, brief, targeted laser pulses are likely cost-optimal for sender and receiver \citep{Benford2010Searching}. Due to highly coherent emission, efficient positional targeting is possible. 
A key challenge of using laser beacons is the contrast problem -- and this is where our paper proposes a different solution. Laser beacons installed on planetary surfaces or on orbits around a host star will always have to compete with the host star's bright light. The most commonly identified solution for this contrast challenge is to use ultrashort laser pulses: With pulses compressed to nanosecond to femtoseconds, a laser system can outshine its host stars' light for a brief fraction of a second. Sophisticated optical SETI searches have, in fact, surveyed over a thousand stars for such ultrashort laser pulses \citep[e.g.,][and references therein]{Vidal2026,Acharyya2023}.

Here -- similarly to \citep[][]{Shostak2009} -- we suggest that a resource-constrained sender may, in fact, use a different approach. Ultrashort laser systems capable of compressing pulses to duration of a fs (10$^{-12}$s) are in existence and used in cutting-edge research laboratories. However, these are very complex and finely tuned systems, that utilize a combination of optical parametric oscillators, mode-locked laser systems, and multi-band non-linear amplification systems to achieve pulse compression and amplification in a non-linear media. These systems may not be ideal choices for a very long-duration, automated, scalable experiment that requires robust, simple, low-cost, mass-producible emission sources. Consider that a current cost of a high-intensity (micro-second) laser diode may be as little as $\sim$\$100, a nano-second system's typical cost is around \$10,000, while a femtosecond laser system exceeds \$100,000. Thus, building a robust, scalable, long-duration system may be more readily achieved by combining many short-pulse ($\sim\mu$m) laser emitters rather than utilizing a few high-end ultrashort laser systems.





This, of course, leaves the sender with the contrast challenge unsolved. The sender may opt for a solution where the laser beacon is removed from the host planetary system, in a system akin to the Voyager spacecrafts. As viewed from $\alpha$ Centauri, the voyager sapcecraft are (as of 2026) separated from the Sun by $\sim$2 arcmin$-$easily resolvable with even small amateur telescopes. A system capable of operating for centuries can leave its host system behind far enough to avoid the contrast problem -- and the longevity of the systems is likely expected given the complexity of the SETI search space.

Arguably, a single pulse -- even if detectable -- would be difficult to confirm and reliably interpret. It is likely that interstellar laser beacons would be long-duration, sustained, pulsed emissions, possibly surrounding their planetary system of origin.

Based on the arguments laid out above, we formulate the following hypothesis: \textbf{There are periodic millisecond/microsecond interstellar beacons operating in the local Galactic neighborhood that are detectable with present-day technology.}

\section{Could we have detected milli-second beacons?}

Perhaps surprisingly, our current sky surveys are practically all blind to sub-second duration light pulses: In most cases, exposure times or integration times range from seconds to hours, and milli- or micro-second pulses would remain unrecognized or be lost in the integrated noise of the exposure. Even if the signal-to-noise ratio is high enough for a source to be detectable, if the timescale of its pulse is much shorter than the integration time of a survey, it would be seen as a faint continuous source and its true nature would likely escape detection.

Over the past three decades, the focus of optical SETI searches has been on targeted (non-parallel) searches of ultrashort laser pulses ($\sim$ns), i.e., mostly star-by-star observations with sensitive avalanche photo-diodes (APDs). While these searches are sensitive to short pulses, by their targeted nature (focusing on stars) they are necessarily blind to interstellar beacons. Due to their design, most such targeted surveys would be highly inefficient in searching for interstellar beacons.

A  notable exception is the PANOSETI concept 
\citep{Wright2018}, which aims to carry out \textit{an-all sky search} for pulsed visible-near-infrared ($\lambda$=350--1650 nm) emission using APDs. A key challenge for this survey is a low-cost but sensitive and light-weight all-sky optical imaging system. The PANOSETI project aims to utilize D=0.46m diameter refractive Fresnel lenses made of acrylic. Each lens would have an approximately 100 deg$^2$ field of view covered by 32$\times$32 pixels (APDs). A proof of concept demonstration was successful \citep{Maire2019} and showed sensitivity to nanosecond pulses. To our knowledge, this exciting concept has not been deployed at full scale. 
A key strength of this survey is its ability to carry out a very wide field of view search for nano-second pulses. A limitation is the very large beam size (resolution elements are $\sim10^{\circ}$), which may lead to severe signal dilution and may limit sensitivity. 

\section{The Interstellar Beacons Project}

Compelled by the arguments laid out above, we propose the \textit{Interstellar Beacons Project} to test the hypothesis that micro/milli-second interstellar laser beacons are observable to us in the night sky.
This hypothesis is testable by \textbf{scanning the entire sky for interstellar laser beacons down to $\tau$=1 ms (threshold) to 1 microsecond (goal) pulse durations}.  Here, we propose a search strategy with a very wide field of view but also high spatial (5''$\times$5'' pixels) and temporal resolution ($\sim$10$\mu$s).
Utilizing a high pixel count (32 mega-pixel) detector allows for a highly parallelized search and yields a $10^6-10^8$ efficiency gain in comparison to existing survey strategies. At the same time, the fine resolution enhances sensitivity (by reducing background by a factor of $~10^4$) and allowing for source localization (even for moving objects $-$ Voyager 1 as seen from $\alpha$ Centauri moves by $\sim$10 mas/day relative to the Sun).

This survey strategy has not been possible in the past, but it opened up through the introduction of three new technologies: 
(1) \textit{At-scale data processing:} With advances in edge processing, real-time processing of micro-second cadence imaging data is now within reach with off-the-shelf computing equipment (GPU-clusters). (2) \textit{Rapid, very low-noise detectors:} The new generation of low-noise, quasi-photon counting CMOS detectors, such as the ORCA-Quest 2 qCMOS detector by Hamamatsu, were successfully tested for astronomical applications \citep[][]{Layden2026}. (3) \textit{High-image quality, ultralight-weight, mass-producible optics:} Multi-order, diffractive-refractive lenses that are replicated via glass pressing are now capable of providing high image quality in a compact, lightweight form factor \citep[][]{Apai2019}. 
The combination of these three new technologies now enables on-the-fly processing of high spatio-temporal, high field-of-view, sensitive images -- making the search for micro/milli-second laser pulses far more compelling than it has been in the past. 

Our hypothesis does not naturally motivate a specific duration for transmission: depending on the sender's strategy, transmission to a specific target system may be episodic and may last for days or millennia. Instead of assuming a sender strategy, we base the survey's observing windows on our own practical reality, envisioning an all-sky survey that can be completed in one decade or less. With this duration, we propose to monitor each field for about one week ($10^{11}\times$ the shortest pulse duration). 
We are currently in the early exploratory design phase of the project. We envision the use of two observing stations or mobile observatories for scanning the northern and southern skies simultaneously. Each station would be equipped with 1\,m diameter MODE-lens based telescopes and $\sim$50 degree$^2$ FOV lenses and pixels of $\sim$5"x5", utilizing 8K $\times$ 4K detectors. MODE lens-based systems are optimal because they allow for ultra-wide FOV and high image quality in a low-cost, optomechanically robust implementation.

\section{Sensitivity Study}

We calculate the detection sensitivity for a pulsed laser beacon under the following assumptions for the laser transmitter.
The adopted parameters are based on the capabilities
of current human laser technology. If such a laser beacon can be constructed
with modern human technology, it is reasonable to consider it a practical
signal that could be produced by an extrasolar civilization somewhat more
technologically advanced than humanity.
The adopted parameters and their corresponding motivations are summarized below.

\begin{itemize}
    \item \textbf{Laser beacon source:}
    \begin{itemize}
        \item \textbf{Central wavelength:} 
        $\lambda = 1064~{\rm nm}$. 
        We adopt this wavelength because it lies in a relatively transparent near-infrared atmospheric window and is therefore practical for ground-based laser transmission and detection. Given this advantage, it was also considered for the Breakthrough Starshot laser-array concept and has been used by JPL's Optical Communications Telescope Laboratory (OCTL) for optical communication with the Psyche spacecraft.

        \item \textbf{Peak laser power and pulse duration:} $P_{\rm laser}=10^{8}~{\rm W}$ and $t_{\rm pulse}=10~\mu{\rm s}$. Together, these assumptions correspond to an emitted laser energy of $E_{\rm pulse}=1~{\rm kJ}$ per pulse.  This pulse energy and duration are within the demonstrated capabilities of human laser technology \citep[e.g.,][]{2006JPhy4.133...75K}.
        $D_{\rm tx} = 1~{\rm m}$.
        We adopt a $1~{\rm m}$ laser transmitter diameter based on JPL's Optical Communications Telescope Laboratory (OCTL), which uses its $1~{\rm m}$ telescope to transmit a $1064~{\rm nm}$ uplink laser signal to the Deep Space Optical Communications (DSOC) instrument aboard the Psyche spacecraft.

        \item \textbf{Beam divergence:}
        $\theta_{\rm div}=3~\mu{\rm rad}$.
        This value is based on the full-angle diffraction-limited beam
        divergence of a $D_{\rm tx}=1~{\rm m}$ transmitter operating at
        $\lambda=1064~{\rm nm}$, for which
        $\theta_{\rm div}\simeq
        2\times(1.22\lambda/D_{\rm tx})\simeq2.6~\mu{\rm rad}$.
        We adopt the rounded value of $3~\mu{\rm rad}$ to calculate the
        laser-beam footprint area at a distance $d$ from the transmitter,
        \begin{equation}
        \label{eq:Abeam}
            A_{\rm beam}
            =
            \pi
            \left[
            d\tan\left(\frac{\theta_{\rm div}}{2}\right)
            \right]^2.
        \end{equation}

        \item \textbf{Beam delivery efficiency:}
        We examine both an idealized beam-delivery case,
        $\eta_{\rm beam}=1.0$, and a more conservative case,
        $\eta_{\rm beam}=0.8$, in which $20\%$ of the emitted laser energy
        is lost due to imperfect beam collimation or pointing.

      \end{itemize}
      \item \textbf{The observation setup for searching for such a laser beacon is summarized below:} 
      \begin{itemize} 
        
        \item \textbf{Exposure time/cadence:} $t_{\rm exp}=10~{\mu s}$. 
        
        \item \textbf{Filter:} $Y$-band, with an approximate wavelength coverage of $970$--$1070~{\rm nm}$. This filter includes the adopted laser wavelength of $1064~{\rm nm}$. 
        
        \item \textbf{Sky background:} $m_{\rm sky}=18~{\rm mag~arcsec^{-2}}$ in the $Y$~band \citep{2006MNRAS.367..454H}. 
        
        \item \textbf{Signal-extraction aperture:} $\Omega_{\rm ap}=100~{\rm arcsec^{2}}$. It comes from a $2\times2$ extraction aperture with the pixel sacle of 5~${\rm arcsec/pixel}$. We use this aperture area to calculate the sky-background contribution included in the extracted laser signal. 
        
        \item \textbf{Detector throughput:} $\eta_{\rm det}=0.9$. We assume that $90\%$ of the laser and sky-background photons reaching the detector are converted into detected photoelectrons. This value represents a high-efficiency detector assumption at the adopted laser wavelength.

        \item \textbf{Atmospheric transmission of the laser signal:} $T_{\rm atm}=0.9$. 
        We assume that $90\%$ of the incoming laser photons of $1064~{\rm nm}$ are transmitted through the Earth's atmosphere under favorable weather and low-airmass observing conditions at a high-altitude site of approximately $2$--$3~{\rm km}$ elevation \citep{BiswasPiazzolla2006}.
      
        \item \textbf{Read noise:} $\sigma_{\rm read}=1~{\rm e^{-}~pixel^{-1}~read^{-1}}$. This value is motivated by the performance of modern scientific CMOS detectors. 


        \item \textbf{Additional assumptions:} We do not include contamination from background stellar sources or attenuation due to interstellar extinction in the present sensitivity calculation. 
      
      \end{itemize}
\end{itemize}

We assume that usable observations are obtained over $40\%$ of a one-week
observing interval. If the laser beacon emits $N_{\rm pulse}=10^{8}$ pulses
during this effective observing time, the corresponding pulse period is
approximately $2.4~{\rm ms}$ or a pulse frequency $f_{pulse}$=413~Hz. 
In practice, the pulse period would not be
known in advance. Therefore, post-processing and periodogram/frequency analysis, or similar methods (such as cross-correlation analysis over trial pulse
periods or grid search) would be required to search for a possible periodic laser-pulse
signal in the observed light curve. Although the signal from an individual
pulse may be too weak to detect, the detection significance can be enhanced
once a possible periodic signal is identified by aligning and stacking the
corresponding on-pulse exposure windows.

In our calculation, both the adopted pulse duration and the exposure time are
$t_{\rm pulse}=t_{\rm exp}=10~\mu{\rm s}$. Therefore, each selected on-pulse
exposure window contains one complete laser pulse. For
$N_{\rm pulse}=10^{8}$ pulses, the total integrated laser-signal time is
$N_{\rm pulse}t_{\rm pulse}=1000~{\rm s}$. Since the exposure time is matched
to the pulse duration, the total selected exposure time is also
$N_{\rm pulse}t_{\rm exp}=1000~{\rm s}$. The stacked signal can therefore be
treated as an integrated $1000~{\rm s}$ on-pulse signal, while the read-noise
variance is accumulated over all individual exposures.

We estimate the signal-to-noise ratio (SNR) after integrating the signal contribution from $N_{\rm pulse}$ pulses as
\begin{equation}
{\rm SNR}
=
\frac{N_{\rm sig,tot}}
{\left(
N_{\rm sig,tot}
+
N_{\rm sky,tot}
+
N_{\rm dark,tot}
+
\sigma_{\rm read,tot}^{2}
\right)^{1/2}},
\label{eq:laser_snr}
\end{equation}
where \begin{align} N_{\rm sig,tot} &= N_{\rm pulse}N_{\rm sig,1}, \\[3pt] N_{\rm sky,tot} &= N_{\rm pulse}N_{\rm sky,1} \nonumber \\ &= N_{\rm pulse} \eta_{\rm det} \frac{ F_{\lambda,{\rm sky}} \Delta\lambda \Omega_{\rm ap} A_{\rm tel} t_{\rm exp} }{ hc/\lambda }, \\[3pt] N_{\rm dark,tot} &= N_{\rm pulse}N_{\rm dark,1}, \\[3pt] \sigma_{\rm read,tot}^{2} &= N_{\rm pulse}\sigma_{\rm read,1}^{2}. \end{align}
Here, the subscript 1 denotes the contribution from a single pulse.
$F_{\lambda,{\rm sky}}$ is the sky-background flux density corresponding to $m_{\rm sky}=18~{\rm AB~mag~arcsec^{-2}}$, $\Delta\lambda = 100~\rm nm$ is the filter bandwidth, $A_{\rm tel}=\pi(D_{\rm tel}/2)^2$,  and $\Omega_{\rm ap}=100~{\rm arcsec^{2}}$ is the signal-extraction aperture. 
We neglect the contribution from dark current, i.e. $N_{\rm dark,tot}=0$, since the adopted exposure time for each pulse is only $10~\mu{\rm s}$, the accumulated dark-current counts are negligible compared with the other noise terms for a cooled detector.
The detected laser-signal counts in one selected on-pulse exposure are
\begin{equation}
N_{\rm sig,1}
=
\eta_{\rm det}
T_{\rm atm}
\eta_{\rm beam}
\left(\frac{A_{\rm tel}}{A_{\rm beam}}\right)
\frac{P_{\rm laser} t_{\rm sig}}{hc/\lambda},
\label{eq:laser_signal_counts}
\end{equation}
where $t_{\rm sig}=10~\mu{\rm s}$ and $A_{\rm beam}$ is the laser-beam footprint area at the observer, as defined in Equation~\ref{eq:Abeam}.

\begin{figure*}[ht!]
\centering
\includegraphics[width=1.0\textwidth, height=0.8\textheight, keepaspectratio]{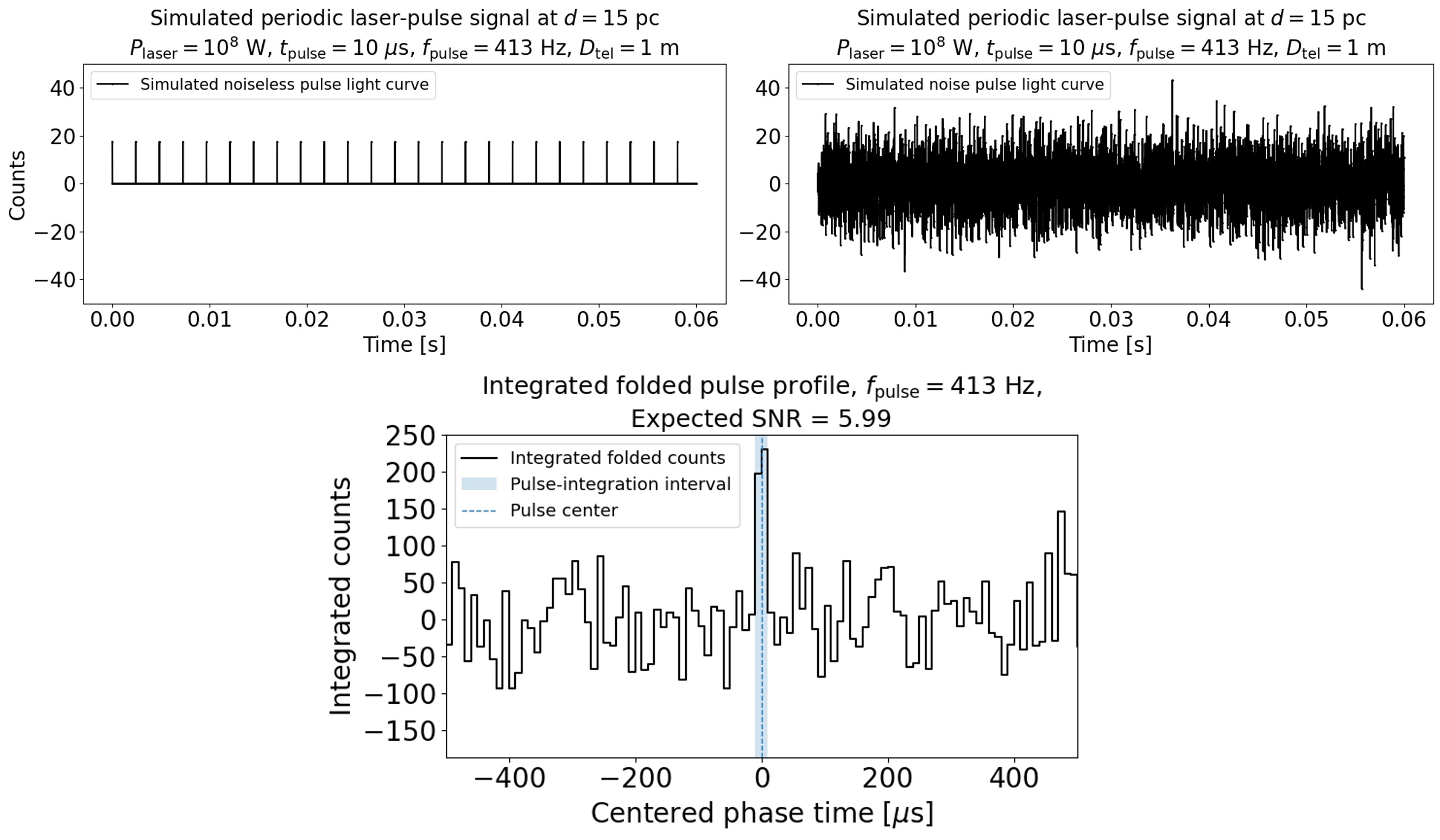}
\caption{Simulated light curve for a periodic laser beacon at $f_{\rm pulse}=413~{\rm Hz}$ at $d=15~{\rm pc}$ observed with a $D_{\rm tel}=1~{\rm m}$ telescope at a cadence of $10~\mu{\rm s}$. The upper panels show a short segment of the noiseless pulse signal (left) and the corresponding signal including the estimated noise contributions (right). The bottom panel shows the integrated folded pulse profile after combining $10^{8}$ pulses, resulting in an expected ${\rm SNR}$ of approximately $5.99$.}
\label{fig:simulated_laser_light_curve}
\end{figure*}

Figure~\ref{fig:simulated_laser_light_curve} shows an example of the simulated light curve for a laser beacon with the parameters described above, located at $d=15~{\rm pc}$ and observed with a $D_{\rm tel}=1~{\rm m}$ telescope at a cadence of $10~\mu{\rm s}$. 
The upper-left panel shows the noiseless periodic laser-pulse signal with a pulse frequency of $f_{\rm pulse}=413~{\rm Hz}$. The upper-right panel shows the corresponding simulated light curve after including the estimated noise contributions. In this light curve, the individual laser pulses cannot be visually identified above the noise. The bottom panel shows the integrated folded pulse profile after recovering the candidate pulse frequency and combining the signal from $10^{8}$ pulses during the one-week observing interval. 
Although the individual pulses are not detectable in the noisy light curve, their periodic signal becomes detectable after integration, resulting in an expected detection significance of ${\rm SNR}=5.99$.

\begin{figure*}[ht!]
\centering
\includegraphics[width=0.6\textwidth, height=0.5\textheight, keepaspectratio]{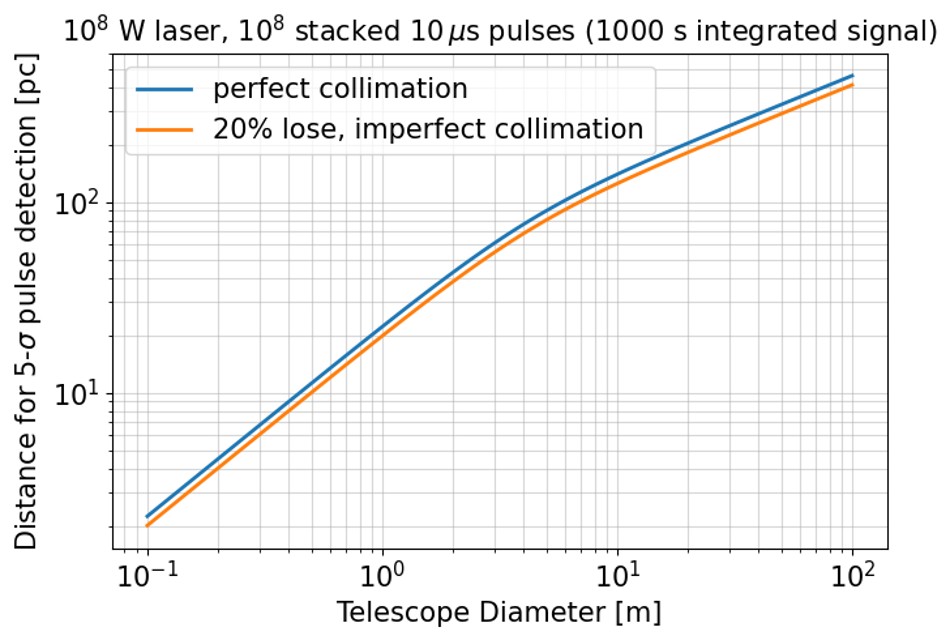}
\caption{Maximum distance for a $5\sigma$ detection of a periodic pulsed laser beacon as a function of telescope diameter. The calculation assumes a peak laser power of $10^{8}~{\rm W}$ and the integration of $10^{8}$ pulses with a pulse duration of $10~\mu{\rm s}$, corresponding to a total integrated laser-signal time of $1000~{\rm s}$. The blue curve represents the idealized perfect-collimation case, while the orange curve assumes a $20\%$ loss of emitted laser energy due to imperfect collimation.}
\label{fig:laser_detection_distance}
\end{figure*}

Figure~\ref{fig:laser_detection_distance} shows the result of our sensitivity calculation: the maximum distance at which the assumed pulsed laser beacon can be detected at a significance of at least $5\sigma$ as a function of the telescope diameter. 
Two beam-delivery scenarios are shown in the figure: an idealized case with perfect collimation (blue) and a more conservative case in which $20\%$ of the emitted laser energy is lost due to imperfect collimation (orange). For $D_{\rm tel}=1~{\rm m}$, such a laser beacon could be detected out to a distance of approximately $20$--$22~{\rm pc}$, depending on the assumed beam-delivery efficiency.
Given that the current full-sky census within $20~{\rm pc}$ of the Sun contains approximately $3000$ A- to M-type stars \citep{2024ApJS..271...55K}, this sensitivity range would allow us to search for interstellar laser-beacon signals sent by exosolar civilizations from these nearby stars in the solar neighborhood.




\section{Summary}

In this paper, we argued that a \textit{resource-limited} civilization's approach for interstellar communication would differ from approaches not constrained by resources. Specifically, we argue that resource limits will compel civilizations to maximize the chance of discovery at low cost and over long project durations. Thus, senders will build systems that are easy-to-locate, long-lived, and autonomous. 
These considerations suggest that ultrashort laser pulses -- the primary assumed modus of optical communication -- may not, in fact, be the optimal solution. 

Instead, we propose that there may be an abundance of replicated, interstellar laser beacons that produce longer-duration (ms to $\mu$s) pulses. Beacons may be within planetary systems or interstellar (very weakly or not bound to stars). Such interstellar laser beacons do not suffer from contrast challenges and do not need to be ultrashort pulsed to be detectable. Thus, they may be optimal considering transmission/reception economics.

Our hypothesis predicts that there may be a large number of pulsed laser emitters that could be visible to us right now -- but our current
surveys are practically blind to such sources (along with the rest of the milli-second universe). 
Although interstellar laser beacons may be key to efficient communication strategy, we still lack at-scale deployable pulsed beacon sensitivity. We showed that three emerging new technologies now open a window for relatively low-cost experiments. Progress thanks to Moore’s Law (impacting both processing and sensing) in combination with new optical technology (MODE lenses) offers opportunity for low-cost realization of an all-sky survey exploring the micro/milli-second sky. 

Finally, the survey proposed is not only compelling for SETI studies but it has an exciting potential for astrophysical studies \citep[][]{Sheikh2020}, opening up the yet unseen  milli-second near-infrared/visible light universe for exploration.

\bibliographystyle{iaulike}
\bibliography{sample}
\end{document}